# NONEQUILIBRIUM THERMODYNAMICS WITH THERMODYNAMIC PARAMETER OF LIFETIME OF SYSTEM. I.

V. V. Ryazanov

Institute for Nuclear Research, Kiev, Ukraine, e-mail: vryazan19@gmail.com

**Abstrac**t

To describe the nonequilibrium states of the system, a new thermodynamic parameter - system lifetime - is introduced. Statistical distributions that describe the behavior of energy and lifetime are recorded. Entropy and obtained thermodynamic relations are compared with the results of Extended Irreversible thermodynamics, where as an additional parameter selected fluxes. For the case of thermal conductivity explicit expressions are obtained for average lifetime and conjugate thermodynamic quantity. It is shown that near the equilibrium state, when is only one stationary nonequilibrium state, flows can only reduce the average system lifetime. But there are possibilities to increase the average lifetime of the system.

1. **Introduction**

Limitations appear in classical linear nonequilibrium thermodynamics [1] in the description of such phenomena as the propagation and absorption of ultrasound in liquids, the density profile of shock waves in gases, etc. Attempts to overcome these limitations led to the creation of Extended Irreversible thermodynamics [2], in which a variable, additional to the conserved values, the values of the flows are selected. The need to select additional parameters related to the deviation of the system from equilibrium was noted in [3].

Recently stochastic nonequilibrium thermodynamics has been developing [4], where, as in the present work, the finiteness of the observed time interval is taken into account. Informational statistical thermodynamics in [5, 6] was developed based on the method of the nonequilibrium statistical operator (*NSO*) Zubarev [7 8]. The relation between *NSO* and the distributions used in this paper was noted in [9]. In [10, 11, 12, 13] is introduced a distribution that contains the lifetime of the system as an additional thermodynamic parameter. The distributions for the lifetime are described by the Pontryagin equations [14], the conjugate kinetic equations for the energy distribution of the system, the main variable of statistical thermodynamics. The value of energy occupies a special place among other physical quantities in the statistical description of equilibrium systems. It can be assumed that the lifetime of the system (first-passage time) plays a similar role in the nonequilibrium case.

The lifetime is determined by the evolution of the system and depends on its properties. And vice versa: the properties of the system depend on its lifetime. An important characteristic, for example, of a person is his age. In thermodynamics and statistical physics it is assumed that statistical systems tend to an equilibrium state. We consider open systems for which the equilibrium state is poorly defined. But all real physical systems of finite sizes have a finite lifetime. This is a universal physical characteristic. Many systems degenerate without reaching equilibrium. Lifetime is a representative physical quantity.

The lifetime depends on the energy of the system, energy flows, the size of the system, and other quantities. An exchange with a thermostat occurs, for example, of energy or the number of particles on which the lifetime depends, and not the lifetime. Lifetime is a macroscopic parameter that characterizes the entire system and its interaction with the environment.

The identification of general properties and general statistical laws that are independent of the structure of matter, universal, is the main task of the thermodynamic method of description. The finiteness of the lifetime of all real physical systems is a universal property, and its inclusion in the thermodynamic description seems natural. The lifetime is an observable, compact, well-defined and physically understandable quantity that reflects many important features of the behavior of the system.

In [3], it was noted that the nonequilibrium state is characterized by an additional macroscopic parameter in the description of the system. In the method of the nonequilibrium statistical operator (*NSO*) [7]-[8] this is the time $t-t_0$ elapsed from the birth of the system, the time of the past life, the time until the first crossing of the zero level in the inverse time, the age of the system [9]. The *NSO* was obtained in [7] by averaging over the initial time; in [9] it was shown that this corresponds to averaging over the distribution of the system lifetime. In [15] it was noted that the state of the system at the current moment of time depends on the entire previous evolution of nonequilibrium processes developing in it and, accordingly, on the time of the past life of the system, its age. Therefore, we choose the random lifetime of the system as the thermodynamic parameter.

Real systems have a finite lifetime, which affects their properties and the properties of their environment. The lifetime of a system is a fundamental quantity that has a dual nature, connected both with the flow of external time and with the properties of the system.

The distribution with the lifetime contains two different time scales: the first for the energy of the system *u* and the second for the lifetime *Γ*, which takes into account long-term correlations and long-term changes in *u* using the thermodynamically conjugated to lifetime value *γ*. A similar operation is contained in the *NSO*. An analogy can be drawn between the approaches of Prigozhine [16] and Zubarev [7]-[8], in the latter the dynamics of the behavior of the system and the contribution of correlations throughout the past lifetime of the system are taken into account.

When considering systems of finite sizes, it is essential to take into account the finiteness of their lifetime. In expressions for kinetic coefficients, kinetic equations, entropy production, generalized transport equations, average fluxes, for all nonequilibrium effects, more accurate dissipative effects proportional to the inverse average lifetime can be taken into account [11]. If the lifetime increases indefinitely (for example, with an unlimited increase in the size of the system), these additives disappear. The source in the Liouville equation is related to the fact that a normally functioning system is in a stationary nonequilibrium state, characterized by a given deviation from equilibrium and the production of entropy. Each state of the system corresponds to its specific lifetime, associated with the magnitude of the flows and sources in the system and its deviation from equilibrium. The effects that the system is exposed to when interacting with the environment cause deviations from the normal steady state and dissipative effects, change the degree of removal from the stationary state, the entropy of the system, and its lifetime. The expressions for the average lifetime depend on the exchange with the environment, the interaction of the system and the environment.

An open system is considered, dynamic quantities under the influence of interaction with the environment become random; a macroscopic description is carried out. The lifetime is introduced as a phenomenological macroscopic quantity. The lifetime is not a dynamic quantity (although it depends on the phase coordinates). This is a statistical value with known equations for the probability density distribution of the lifetime. The introduction of a lifetime as a fundamental thermodynamic random variable seems to be an important and necessary element for the correct description of nonequilibrium phenomena.

The choice of the lifetime as a thermodynamic parameter is possible by determining the thermodynamic parameters. For example, in [3] and [17] it is said that "Any function $B(z)$ of dynamic variables ($z=q_1,...,q_N, p_1,...,p_N$) that is macroscopic in nature is a random internal thermodynamic parameter". The fact that the lifetime (1) $\Gamma(z)$ depends on $z$ can be seen from the equations for the distribution of the lifetime in the Markov model [14].

The beginning and end of the lifetime, the highlighted stopping times, are significant. A random process for a lifetime is a subordinate random process, and a lifetime is a subordinate random physical quantity that depends and is determined by energy, number of particles, flows, size of the system, interactions in it and with the environment. The introduction of the lifetime $\Gamma$ reflects a possible complication of the structure of the phase space, which may contain regions of different behaviors (for example, attractors with anomalously long lifetimes, etc.). It follows from the theory of random processes that the existence and finiteness of the lifetime $\Gamma$ is ensured by the presence of stationary states, which physically corresponds to the existence of stationary structures.

In the present work, as in [10, 11, 12, 13], it is proposed to choose an additional parameter related to the deviation of the system from equilibrium [2, 3], in the form of the system lifetime. The lifetime is defined as the random moment of the first achievement of a certain boundary (1), for example, of the zero level by a stochastic process that describes the macroscopic parameter $y(t)$, where $y(t)$ is the order parameter of the system, for example, its energy, number of particles,

$$\Gamma_x = \inf\{t: y(t)=0\}, y(0)=x>0 . \tag{1}$$

The lifetime is the time period till the moment of the first (random) achievement of a certain level (for example, zero level) by a random process $y(t)$ for the macroscopic parameter (for example, energy or particle number). The lifetime is the slave process in the terminology of the random processes theory [18]; that of $y(t)$ determines the behaviour of lifetime. The lifetime depends on the energy of a system, its size, fluxes of energy. Therefore a system exchanges with thermostat the energy, the particle number, but not the "lifetime". The lifetime $\Gamma$ is a macroparameter characterizing the system and its interaction with the environment. It is an observable, well-defined and physically well understandable quantity reflecting important system peculiarities.

The lifetime of the system is associated with the time of the stable existence of the system, its response to external and internal influences, with the stability and adaptive abilities of the system. We assume that the lifetime is associated with a deviation of the system from equilibrium.

Stratonovich in [19] used the term "lifetime" with respect to of phenomena considered. Following synonyms are encountered by us: the first-passage time (for some given level), escape time, busy period (in the queuing theory), etc. First published paper on the subject is [20] where the Pontryagin equations for the lifetime distribution are obtained. In 1940 the Kramers work [21] appeared which dealt with the escape time from the potential well. These questions are discussed in books by van Kampen [22], Gardiner [23] and other authors [24-30]. The lifetime plays part in the theory of phase transitions, chemical reactions, in investigating dynamics of the complex biomolecules, calculating the coefficient of the surface diffusion in semiconductors, in nuclei, elementary particles, spin glasses, spectroscopy, trap systems, in the theory of metastable states etc.

The termination of the lifetime, functioning of the system, can occur at any moment of a complex hierarchy of different times, corresponding to the laws of behavior of a complex system. The thermodynamic state of a system at a given moment of time is defined as a set of macroscopic parameters that distinguish it from the environment.

The characteristics of the lifetime $\Gamma$ depend on the main process $y(t)$. The introduction of $\Gamma$ means the effective accounting of more information than that contained in the linear terms of the canonical distribution $exp\{-\beta E\}/Z(\beta)$. Real systems have a finite lifetime, which significantly affects their properties and the properties of their environment.

The choice of the lifetime as a thermodynamic parameter can be justified using the *NSO* method [7, 8], which is interpreted in [9] as the averaging of the quasiequilibrium statistical operator $\rho_q$ over the distribution of the (past) lifetime of the system $p_q(m), m=t-t_0$. Then for $p_q(m)=\varepsilon exp\{-\varepsilon m\}$

$$ln\rho(t)=\int_0^\infty p_q(m)ln\rho_q(t-m,m)dm=ln\rho_q(t,0)-\int_0^\infty exp\{-\varepsilon m\}(dln\rho_q(t-m,-m)/dm)dm. \quad (2)$$

where $\rho_q(t,0)$ is a locally equilibrium operator. The entropy production operator [7, 8] is $\sigma(t-m,-m)=dln\rho_q(t-m,-m)/dm$. If $\sigma(t-m,-m)\approx\sigma(t)$, weakly depends on $m=t-t_0$, then equality (2) takes the form $ln\rho(t)=ln\rho_q(t,0)+<\sigma(t)>\varepsilon^{-1}=ln\rho_q(t,0)+<\sigma(t)><\Gamma>$, since in the interpretation of [9] $\varepsilon^{-1}=<\Gamma>=<t-t_0>$ is the average of (past) lifetime.

Note that the same distribution was obtained in [31]. Expression (2) can be rewritten in the form (4) - (5), or, in a more general form, $ln\rho(t)=ln\rho_q(t,0)+ln[Z(\beta)/Z(\beta,\gamma)]-\gamma\Gamma$, where $\rho=exp\{-\beta E-\gamma\Gamma\}/Z(\beta,\gamma)$; $\rho_q=exp\{-\beta E\}/Z(\beta)$ (in (4) instead of $<\Gamma>(dln\rho_q(t-y_1,-y_1)/dt)$ there is a random variable $\Gamma$ and the Lagrange multiplier $\gamma$ in: $\gamma\Gamma+lnZ(\beta,\gamma)/Z(\beta)$). Thus, averaging over the distribution of the lifetime in the Zubarev *NSO* corresponds to the use of a random variable of the lifetime for some approximations — this is a significant difference between the distribution (4) - (5) from (2) and the results of [7, 8, 31].

Section 2 contains distributions describing a system in a stationary nonequilibrium state with a thermodynamic parameter of the lifetime. The assumptions made at the same time are noted.

In the third section, the entropy and its differentials are determined both for the distribution depending on the energy $y=u$ and the lifetime $\Gamma$, when the entropy is equal to $S_\Gamma$ (7), and for the distribution depending only on $y=u$. It is assumed that temporary changes in the entropy of $S_\Gamma$ are caused by flows of entropy entering the system. Then the thermodynamic force $\gamma$ conjugated to the value $\Gamma$ is connected with the flows flowing through the system in a stationary nonequilibrium state. The results are compared with Extended Irreversible thermodynamics [2], where the flux values are chosen as an additional thermodynamic parameter.

The deviation of the nonequilibrium entropy from its equilibrium value (deviations from any fixed value of entropy can also be chosen) are expressed through the production of entropy and the change in the lifetime.

Explicit expressions for the average lifetime $\Gamma$ and the thermodynamic parameter $\gamma$ that is conjugated to $\Gamma$ value are obtained. For one stationary nonequilibrium state and for the exponential distribution for the lifetime, flows can only reduce the average lifetime.

## 2. Distributions for energy and lifetime

The lifetimes are affected by attractors, metastable states, phase transitions, and other physical features of the system that depend on dynamic variables ($z=q_1,...,q_N, p_1,...,p_N$), $q$ and $p$ are the coordinates and momenta of $N$ particles of the system. The lifetime depends on the energy $u$ and the number of particles $N$, depending on $z$. Moreover, the lifetime is a macroscopic and slowly changing quantity.

Gibbs equilibrium distribution for microcanonic probability density in phase space z corresponds to a condition of equiprobability of all possible microstates compatible with the given value of a macroscopical variable. We shall assume, that transition of system in a nonequilibrium condition breaks equality of probabilities, characteristically for an equilibrium case. One introduces additional observable macroparameter, thus extending the phase space (containing additional degenerate, absorbing states). We shall assume validity of a principle of equal probabilities for the extended phase space divided into cells with constant values $(u,\Gamma)$ (instead of phase cells with constant values $u$).

The distribution for the lifetime $\Gamma$ (1) depends on the macroscopic values of $y(t)$. Suppose that the process $y(t)=u$ is the energy of the system (you can also choose the number of particles). The standard procedure (for example, [32]) allows you to rewrite the relation between the distribution density $p(u,\Gamma)=p_{u\Gamma}(x,y)$ and the microscopic (coarse-grained) density $\rho(z;u,\Gamma)$

$$p(u,\Gamma)=\int\delta(u-u(z))\delta(\Gamma-\Gamma(z))\rho(z;u,\Gamma)dz=\rho(u,\Gamma)\omega(u,\Gamma). \qquad (3)$$

The factor $\omega(u)$ is replaced by $\omega(u,\Gamma)$, the volume of the hypersurface in the phase space containing fixed values of $u$ and $\Gamma$. If $\mu(u,\Gamma)$ is the number of states in the phase space with parameters less than $u$ and $\Gamma$, then $\omega(u,\Gamma)=d^2\mu(u,\Gamma)/dud\Gamma$. Moreover, $\int\omega(u,\Gamma)d\Gamma=\omega(u)$. The number of phase points with parameters lying in the interval between $u$, $u+du$; $\Gamma$, $\Gamma+d\Gamma$, is equal to $\omega(u,\Gamma)dud\Gamma$.

We now use the principle of equal probabilities for an extended phase space with cells $(u,\Gamma)$ (in this case, realizations of the space of elementary events change). Using the principle of maximum entropy [33], we write the expression for the microscopic probability density in the extended phase space

$$\rho(z;u,\Gamma)=\exp\{-\beta u(z)-\gamma\Gamma(z)\}/Z(\beta,\gamma), \qquad (4)$$

where

$$Z(\beta,\gamma) = \int e^{-\beta u-\gamma\Gamma}dz = \iint du\ d\Gamma\ \omega(u,\Gamma)e^{-\beta u-\gamma\Gamma} \qquad (5)$$

is the partition function, $\beta$ and $\gamma$ are the Lagrange multipliers satisfying the following expressions for the averages:

$$<u>=-\partial \ln Z/\partial\beta|_\gamma; \qquad <\Gamma>=-\partial \ln Z/\partial\gamma|_\beta. \qquad (6)$$

The thermodynamically conjugated lifetime value $\gamma$, as can be seen from (2), is associated with the production and flows of entropy, which characterize the peculiarities of the nonequilibrium processes in an open statistical system. At $\gamma=0$, the nonequilibrium distribution (4) yield the equilibrium Gibbs distribution. The distribution over a lifetime is a generalization of Gibbs distribution to a nonequilibrium situation. The canonical Gibbs distribution is obtained from the microcanonical ensemble in the zeroth approximation by the interaction of the system with the environment.

One can thus consider (4)-(5) as a generalization of the Gibbs statistics to cover the nonequilibrium situation. Such physical phenomena as the metastability, phase transitions, stationary nonequilibrium states are known to violate the equiprobability of the phase space points. The value $\gamma$ can be regarded as a measure of the deviation from the equiprobability hypothesis. The value of $\gamma = 0$ characterizes equilibrium isolated systems, presenting thus an idealisation. If the detailed balance is satisfied and in equilibrium $\gamma \neq 0$. Mathematically introducing lifetime means acquiring additional information regarding an underlying stochastic process, beyond merely knowledge of its stationary distribution, exploring the (stationary) properties of its slave process.

Using the lifetime $\Gamma$, an effective account is taken of this interaction (similarly to the methods of McLennan [34] and Zubarev *NSO* [7, 8]). Physical phenomena such as metastability, phase transitions, attracting attractors and repellers, stationary nonequilibrium states violate the equal probability hypothesis of phase space points. The value of $\gamma$ can be considered as a measure of deviation from this hypothesis and the Gibbs distribution. Many works are close to nonequilibrium thermodynamics obtained from the distribution with the lifetime as a thermodynamic parameter [10, 11]. For example:

- Extened Irreversible thermodynamics [2], in which flows, quantities inverse to the lifetime are selected as an additional thermodynamic parameter.
- Informational statistical thermodynamics [5, 6], based on the Zubarev *NSO*, in which an infinite number of thermodynamic parameters, flows of all orders are selected. The truncation operation, the limitation by a finite number of parameters, leads to the fact that the lifetime is also determined with some accuracy. It was noted in [5, 6] that the *NSO* contains a hypothesis of a non-mechanical nature. This is a hypothesis about the possibility of using and finiteness of the system lifetime.

Let us underline the principal features of the suggested approach.

1. We introduce a novel variable $\Gamma$ which can be used to derive additional information about a system in the stationary nonequilibrium state. We suppose that $\Gamma$ is a measurable quantity at macroscopic level, thus values like entropy which are related to the order parameter (principal macroscopic variable) can be defined. At the mesoscopic level the variable $\Gamma$ is introduced as a variable with operational characteristics of a random process slave with respect to the process describing the order parameter.

2. We suppose that thermodynamic forces $\gamma$ related to the novel variable can be defined. One can introduce the ``equations of state'' $\beta(<u>,<\Gamma>)$, $\gamma(<u>,<\Gamma>)$ (*u is* energy). Thus we introduce the mapping (at least approximate) of the external restrictions on the point in the plane $\beta, \gamma$.

3. We suppose that a "refined" structure factor $\omega(u,\Gamma)$ can be introduced which satisfies the condition $\int\omega(u,\Gamma)d\Gamma=\omega(u)$ (ordinary structure factor). This function (like $\omega(u)$) is the internal (inherent) property of a system. At the mesoscopic level we can ascribe to this function some inherent to the system (at given restrictions $(\beta_0,\gamma_0)$) random process. The structure factor has the meaning of the joint probability density for the values $u, \Gamma$ understood as the stationary distribution of this process.

4. It is supposed that at least for certain classes of influences the resulting distribution has the form (4), that is the change of the principal random process belongs to some class of the invariance leading to this distribution which explains how one can pass from the process in the reper point $(\beta_0,\gamma_0)$ (for example, in equilibrium when $\gamma=0$ and $\beta=1/k_BT$) to a system in an arbitrary nonequilibrium stationary state. The thermodynamic forces should be chosen so that the distribution lead to new (measurable) values of $(<u>,<\Gamma>)$.

## 3. Generalized thermodynamics of systems with parameter of lifetime

If $\gamma=0$ and $\beta=\beta_0=T^{-1}_{eq}$, where $T_{eq}$ is the equilibrium temperature, expressions (4), (5) go over to the equilibrium Gibbs distribution. Distributions (4), (5) can be considered as a generalization of Gibbs distribution to a nonequilibrium situation. In the general case, the value of $\Gamma$ can be selected as a subprocess in a form different from that described above. Then the equilibrium situation will mean some curve in the $(\beta,\gamma)$ plane, which does not necessarily coincide with the line $\gamma=0$. Features of the nonanalytic behavior of the partition function (or its derivatives) on the $(\beta,\gamma)$ plane indicate nonequilibrium phase transitions.

Assuming the measurability of the lifetime, we introduce the entropy $S_\Gamma$ corresponding to distribution (4) by the relation

$$S_\Gamma=-<\ln\rho(z;u,\Gamma)>=\beta<u>+\gamma<\Gamma>+\ln Z(\beta,\gamma); \qquad dS_\Gamma=\beta d<u>+\gamma d<\Gamma>. \tag{7}$$

For spatially inhomogeneous systems, the quantities $\beta$ and $\gamma$ in the general case depend on the spatial coordinate. Distributions (4) can be considered for a small volume element in which the values of $\beta$ and $\gamma$ are replaced by the average values constant over this volume element over this volume element. In nonequilibrium thermodynamics, the densities of extensive thermodynamic quantities (entropy, internal energy, mass fraction of a component) are considered. We follow this approach, including here and the lifetime.

The Fourier law $q=-\lambda\nabla T$ (where $q$ is the heat flux, $\lambda$ is the heat conductivity coefficient, $T$ is the absolute temperature) of classical nonequilibrium thermodynamics [1] is replaced in Extended Irreversible thermodynamics [2] by the Maxwell-Cattaneo equation of the form

$$q = -\lambda\nabla T - \tau_q\partial q/\partial t, \tag{8}$$

where $\tau_q$ is the correlation time of flows. In this case, the Fourier law is generalized, taking the form:

$$q(t) = -\int_0^t \varphi(t-t`)\nabla T(t`)dt`. \qquad (9)$$

The solution of equation (8) is $q(t) = -\int_0^t [(\lambda/\tau)\exp\{-(t-t`)/\tau\}]\nabla T(t`)dt`$, then the memory function is $\varphi(t-t`) = (\lambda/\tau)\exp\{-(t-t`)/\tau\}$. For $\tau \to 0$, $\varphi(t-t`) = \lambda\delta(t-t`)$. Substituting this value of the memory function in (9), we obtain the Fourier equation.

In distribution (4) - (5), containing the lifetime as a thermodynamic parameter, the joint probability (3) - (4) for the quantities $u$ and $\Gamma$ is

$$P(u,\Gamma) = \frac{e^{-\beta u - \gamma\Gamma}\omega(u,\Gamma)}{Z(\beta,\gamma)}. \qquad (10)$$

Integrating (10) over $\Gamma$, we obtain a distribution of the form

$$P(u) = \int P(u,\Gamma)d\Gamma = \frac{e^{-\beta u}}{Z(\beta,\gamma)}\int_0^\infty e^{-\gamma\Gamma}\omega(u,\Gamma)d\Gamma. \qquad (11)$$

The factor $\omega(u,\Gamma)$ is the joint probability for $u$ and $\Gamma$, considered as the stationary probability of this process. We rewrite the value $\omega(u,\Gamma)$ in the form

$$\omega(u,\Gamma) = \omega(u)\omega_1(u,\Gamma) = \omega(u)\sum_{k=1}^n R_k f_{1k}(\Gamma,u). \qquad (12)$$

In equality (12), it is assumed that there are $n$ classes of states in the system; $R_k$ is the probability that the system is in the $k$-th class of states, $f_{1k}(\Gamma,u)$ is the density of the distribution of the lifetime $\Gamma$ in this class of (ergodic) states (in the general case, $f_{1k}(\Gamma,u)$ depends on $u$). As a physical example of such a situation (characteristic of metals, glasses, etc.), one can cite the potential of many complex physical systems.

Such a situation was considered in [35, 36]. The minimum points of the potential correspond to metastable phases, disordered structures, etc. The phase space of these systems is divided into isolated regions, each of which corresponds to a metastable thermodynamic state, and the number of these regions increases exponentially with increasing total number of (quasi)particles [37].

The finiteness of the lifetime of a system implies the finiteness of its size. The thermodynamic limit is not calculated, although the dimensions of the system are assumed to be so large that it is possible to implement a probabilistic ensemble as the state of the system, and to obtain average values, you can choose not infinite, but very large values of the volume and number of particles with a small error. In the theory of random processes, when the definition (1) is valid, there is the concept of the limit of a sequence of random variables. There are at least four ways to determine this limit. Queuing theory addresses the limits of intervals between receipts and other events.

You can consider finite systems, or you can make the limit transition, as in the theory of random processes. Moreover, the volume of the system can be finite, an infinite number of observations requires infinite time. Infinite time also appears in the definitions of ergodicity and mean equilibrium value. One of the arguments for introducing the thermodynamic limit is the elimination of boundary effects (for example, [38]). But for the finiteness of the lifetime, the existence of a boundary is necessary. In [17], large values of the number of particles $N$ are considered, but there is no transition to the limit.

To obtain an explicit form of the distribution in (12), we use the general results of the mathematical theory of phase enlargement of complex systems [39], from which the following exponential form of the density of the distribution of the lifetime for an enlarged random process follows (see also [19]):

$$p(\Gamma < y) = \Gamma_0^{-1}\exp\{-y/\Gamma_0\} \qquad (13)$$

for one class of stable states, and Erlang distribution density

$$p(\Gamma < y) = \sum_{k=1}^{n} R_k \Gamma_{0k}^{-1} \exp\{-y/\Gamma_{0k}\}; \qquad \sum_{k=1}^{n} R_k = 1 \qquad (14)$$

in the case of several (*n*) classes of ergodic stable states, $\langle \Gamma_{\gamma k} \rangle$ is the average lifetime in the *k*-th class of metastable states with a perturbation $\gamma = \gamma_k$, $\Gamma_{0k} = \langle \Gamma_{0k} \rangle$ is the average lifetime in the *k*-th class of metastable states without a perturbation $\gamma = \gamma_k = 0$, in some basic stationary state.

The exponential form of distribution (13) can be justified using, for example, the principle of maximum entropy, general approaches of statistical physics, and also using approaches of information geometry [40].

The values $\langle \Gamma_{0k} \rangle$ and $\langle \Gamma_{\gamma k} \rangle$ are the averaging of the residence time and degeneracy probabilities over stationary ergodic distributions (in our case, the Gibbs distribution). The physical reason for the implementation of the distribution in the form of (13)-(14) is the presence of weak ergodicity in the system, which in the limit leads to stationary distributions. The mixing of system states at large times leads to distributions (13) - (14). The structural factor $\omega(u,\Gamma)$ has the value of the joint probability density of the values of *u*, $\Gamma$. For distributions (13), (14), the functions $\omega(u,\Gamma)$ from (12) have the form:

$$\omega(u,\Gamma) = \omega(u) \sum_{k=1}^{n} R_k \Gamma_{0k}^{-1} \exp\{-\Gamma/\Gamma_{0k}\}; \quad \omega(u,\Gamma) = \omega(u) \Gamma_0^{-1} \exp\{-\Gamma/\Gamma_0\}; \quad n=1. \qquad (15)$$

The form of the function $p_\Gamma(y)$ (13)-(14) reflects not only the internal properties of the system, but also the influence of the environment on the open system, and the features of its interaction with the environment. A physical interpretation of the exponential distribution for the function $p_\Gamma(u)$ is given in [7, 8]: the system evolves as an isolated system controlled by the Liouville operator. In addition, the system undergoes random transitions, and the phase point representing the system switches from one trajectory to another with an exponential probability under the influence of a "thermostat". Exponential distribution describes completely random systems. The influence of the environment on the system can also be organized in nature, for example, this applies to systems in a nonequilibrium state with input and output unsteady flows. The nature of the interaction with the environment may also change; therefore, various forms of the function $p_\Gamma(y)$ can be used [11].

Note that a value similar to $\gamma$ is determined in [41]–[43] for a fractal repeller object. It is equal to zero for a closed system, and for an open system it is equal to $\Sigma\lambda_i - \lambda_{KS}$, where $\lambda_i$ are Lyapunov exponents, $\lambda_{KS}$ is the Kolmogorov-Sinai entropy. The thermodynamic interpretation of $\gamma$ is given below.

We consider the case when the average lifetime $\Gamma_0$ in (13) depends on a random variable *u*. Then, integrating over $\Gamma$, as in (11), we obtain for *n*=1 from (5), (13), (15) that

$$Z(\beta,\gamma) = \int e^{-\beta u} \omega(u) \frac{du}{1 + \gamma \Gamma_0(u)}. \qquad (16)$$

In Extended Irreversible thermodynamics [2], the differential of the entropy density for the case of thermal conductivity is

$$ds = \theta^{-1} du_\beta - \frac{\tau}{\rho \lambda \theta^2} q dq, \quad s = S/V, \qquad (17)$$

where *s* is the entropy density, *q* is the heat flux, $\theta^{-1}$ is the nonequilibrium temperature, $\rho$ is the mass density, $\lambda$ is the thermal conductivity, $\tau = \tau_q$ is the correlation time of the fluxes from equation (14),

$$s_{eq} = s_{\Gamma|\gamma=0} = \beta u_\beta + \ln Z_\beta, \quad Z_\beta = \int e^{-\beta u} \omega(u) du, \quad u_\beta = -\partial \ln Z_\beta / \partial \beta. \qquad (18)$$

The nonequilibrium temperature is

$$\theta^{-1} = T^{-1} - \frac{1}{2}\frac{\partial}{\partial u_\beta}(\frac{\tau}{\rho\lambda\theta^2})q^2, \tag{19}$$

Entropy and entropy production are equal

$$s(u,q) = s_{eq}(u) - \frac{1}{2}\frac{\tau}{\rho\lambda\theta^2}q^2, \qquad \sigma^s = \frac{1}{\lambda\theta^2}qq. \tag{20}$$

Comparing the expressions for entropy (20) and the differential of entropy (17) from Extended Irreversible thermodynamics with variable values of energy and flows with the same expressions, which include the lifetime obtained from expressions (7), (6), we write the relation

$$ds = \theta^{-1}du_\beta - a_\beta qdq = \beta d\bar{u} + \gamma d\bar{\Gamma} = \beta\frac{\partial \bar{u}}{\partial \beta}\bigg|_\Gamma d\beta + (\gamma + \beta\frac{\partial \bar{u}}{\partial \bar{\Gamma}}\bigg|_\beta)d\bar{\Gamma}, \qquad a_\beta = \frac{\tau}{\rho\lambda\theta^2}, \tag{21}$$

where the variables $\beta$ and $\Gamma$ are selected on the right-hand side. We rewrite (21) in the form

$$\gamma + \beta\frac{\partial \bar{u}}{\partial \bar{\Gamma}}\bigg|_\beta + \beta\frac{\partial \bar{u}}{\partial \beta}\bigg|_\Gamma \frac{\partial \beta}{\partial \bar{\Gamma}}\bigg| = \theta^{-1}\frac{\partial u_\beta}{\partial \bar{\Gamma}}\bigg| - a_\beta q\frac{\partial q}{\partial \bar{\Gamma}}\bigg|. \tag{22}$$

Given a constant value of $\gamma$ in relation (22), we obtain

$$\gamma = \frac{1}{\frac{\partial \bar{\Gamma}}{\partial \beta}\bigg|_\gamma}(\theta^{-1}\frac{\partial u_\beta}{\partial \beta}\bigg|_\gamma - \beta\frac{\partial \bar{u}}{\partial \beta}\bigg|_\gamma). \tag{23}$$

Expression (23) is also written from (22) and when a constant value of $q$ is set. Setting constant values of $\beta$ or $u_\beta$ in (22), we obtain

$$\gamma = -\frac{1}{\frac{\partial \bar{\Gamma}}{\partial \gamma}\bigg|_\beta}(\beta\frac{\partial \bar{u}}{\partial \gamma}\bigg|_\beta + a_\beta q\frac{\partial q}{\partial \gamma}\bigg|_\beta). \tag{24}$$

From (24) and (23) we obtain

$$\frac{\partial q}{\partial \gamma}\bigg|_\beta = \frac{1}{a_\beta q}(-\gamma\frac{\partial \bar{\Gamma}}{\partial \gamma}\bigg|_\beta - \beta\frac{\partial \bar{u}}{\partial \gamma}\bigg|_\beta), \qquad \beta\frac{\partial \bar{u}}{\partial \beta}\bigg|_\gamma = \theta^{-1}\frac{\partial u_\beta}{\partial \beta}\bigg|_\gamma - \gamma\frac{\partial \bar{\Gamma}}{\partial \beta}\bigg|_\gamma. \tag{25}$$

For the time derivative of the average lifetime, expressions are written

$$\gamma\frac{d\bar{\Gamma}}{dt} = \theta^{-1}\frac{du_\beta}{dt} - a_\beta q\frac{dq}{dt} - \beta\frac{d\bar{u}}{dt};$$

$$(\gamma + \beta\frac{\partial \bar{u}}{\partial \bar{\Gamma}}\bigg|_\beta)\frac{d\bar{\Gamma}}{dt} = \theta^{-1}\frac{du_\beta}{dt} - a_\beta q\frac{dq}{dt} - \beta\frac{\partial \bar{u}}{\partial \beta}\bigg|_{\bar{\Gamma}}\frac{d\beta}{dt}; \tag{26}$$

$$(\gamma + \beta\frac{\partial \bar{u}}{\partial \bar{\Gamma}}\bigg|_\gamma)\frac{d\bar{\Gamma}}{dt} = \theta^{-1}\frac{du_\beta}{dt} - a_\beta q\frac{dq}{dt} - \beta\frac{\partial \bar{u}}{\partial \gamma}\bigg|_\Gamma\frac{d\gamma}{dt};$$

$$(\gamma + \beta\frac{\partial \bar{u}}{\partial \bar{\Gamma}}\bigg|_q)\frac{d\bar{\Gamma}}{dt} = \theta^{-1}\frac{du_\beta}{dt} - a_\beta q\frac{dq}{dt} - (\beta\frac{\partial \bar{u}}{\partial q}\bigg|_{\bar{\Gamma}} + \gamma\frac{\partial \bar{\Gamma}}{\partial q}\bigg|_{\bar{\Gamma}})\frac{dq}{dt}; \tag{27}$$

$$(\gamma + \beta\frac{\partial \bar{u}}{\partial \bar{\Gamma}}\bigg|_{u_0})\frac{d\bar{\Gamma}}{dt} = \theta^{-1}\frac{du_\beta}{dt} - a_\beta q\frac{dq}{dt} - (\beta\frac{\partial \bar{u}}{\partial u_0}\bigg|_{\bar{\Gamma}} + \gamma\frac{\partial \bar{\Gamma}}{\partial u_0}\bigg|_{\bar{\Gamma}})\frac{du_\beta}{dt}.$$

Using expressions (6), (16), we write the value $\gamma\langle\Gamma\rangle$ in the relation for entropy (7) in the form:

$$s_\Gamma = -\langle \ln\rho(z;u,\Gamma)\rangle = \beta\langle u\rangle + \gamma\langle\Gamma\rangle + \ln Z(\beta,\gamma) =$$
$$\int e^{-\beta u}\omega(u)(\frac{x}{(1+x)^2})du/Z + \beta\langle u\rangle + \ln Z(\beta,\gamma), \quad x = \gamma\Gamma_0(u), \tag{28}$$

and we equate the quantity $s_\gamma = -\Delta = s_\Gamma - s_{eq}$ and (18) to the corresponding expression of the Extended Irreversible thermodynamics (20) [2], where this value is equal $-a_\beta q^2/2$, $a_\beta = \tau/\rho\lambda T^2$. The value $Z$ is equal to (16). And for the average lifetime in (28), the expression obtained from (16) and (6) is used

$$<\Gamma> = -\frac{\partial \ln Z}{\partial \gamma}\bigg|_\beta = \frac{1}{Z}\int e^{-\beta u}\omega(u)\frac{\Gamma_0 du}{(1+x)^2} . \tag{29}$$

We rewrite expressions (28), (29) in the form

$$Z\gamma\overline{\Gamma} = \int e^{-\beta u}\omega(u)(\frac{x}{(1+x)^2})du = Z[\beta(u_\beta - \overline{u}) - a_\beta q^2/2 - \ln Z(\beta,\gamma)/Z_\beta] . \tag{30}$$

We assume that the fluxes $q$ weakly depend on $\beta$ and neglect this dependence (also in [2]). The internal energy $u_\beta$ does not depend on $\gamma$ and $q$. Thus, in this approximation, the fluxes $q$ depend only on $\gamma$, and $u_\beta$ depends only on $\beta$.

The initial relation (28) has the form

$$s = s_{eq} - a_\beta q^2/2 = \gamma\overline{\Gamma} + \beta\overline{u} + \ln Z = \beta u_\beta + \ln Z_\beta - a_\beta q^2/2 . \tag{31}$$

From relations (17) and (6) we obtain (29) and

$$<u> = -\frac{\partial \ln Z}{\partial \beta}\bigg|_\gamma = \frac{1}{Z}\int e^{-\beta u}\omega(u)[\frac{u}{1+\gamma\Gamma_0(u)} + \frac{d(\gamma\Gamma_0)/d\beta}{(1+\gamma\Gamma_0(u))^2}]du . \tag{32}$$

We substitute the expressions (29) and (32) in (31), multiply by $Z$ and equate the integrands. We get the expression

$$\beta\frac{dx}{d\beta} + x[1 - \beta(u_\beta - u) + \ln Z/Z_\beta + a_\beta q^2/2] - \beta(u_\beta - u) + \ln Z/Z_\beta + a_\beta q^2/2 = 0 . \tag{33}$$

Differentiating (31) with respect to $\beta$, we obtain

$$\gamma\frac{\partial\overline{\Gamma}}{\partial\beta} + \beta\frac{\partial(\overline{u} - u_\beta)}{\partial\beta} + \frac{1}{2}q^2\frac{\partial a_\beta}{\partial\beta} = 0 . \tag{34}$$

If now in (31) we take into account dependence (16) only in $lnZ$, differentiate with respect to $\beta$, multiply by $Z$ and equate the integrands, then for $dx/d\beta$, taking into account (34), we obtain

$$\beta\frac{dx}{d\beta} - x\beta(\overline{u} - u) - \beta(\overline{u} - u) = 0 . \tag{35}$$

Comparing this expression (35) with (33), we obtain

$$x = \frac{a_3}{1 - a_3}, \quad a_3 = \beta(u_\beta - \overline{u}) - \ln Z/Z_\beta - a_\beta q^2/2 . \tag{36}$$

It can be seen that the right-hand side of (36) is independent of random variables. Therefore, averaging applies only to the left side of expression (36). Averaged over the equilibrium distribution and below we mean by the value $x = \gamma\overline{\Gamma}_0(u)$ this value averaged over the equilibrium distribution. We do not use the nonequilibrium distribution (4), since $\Gamma_0(u)$ it does not depend on $\gamma$ and $\Gamma$. We assume that the average value is independent of $\gamma$. Substituting this expression (36) into (16), we obtain, taking into account (31), that

$$Z = Z_\beta[1 - \beta(u_\beta - \overline{u}) + \ln Z/Z_\beta + a_\beta q^2/2] = Z_\beta[1 - \gamma\overline{\Gamma}] , \tag{37}$$

i.e., relation (40).

Differentiating (37) with respect to $\gamma$, we obtain, taking into account (6)

$$\gamma\langle\Gamma\rangle^2 + \gamma\frac{\partial\langle\Gamma\rangle}{\partial\gamma} = 0 . \tag{38}$$

From this differential equation with the initial condition $\Gamma_{\gamma=0}=\Gamma_0$ we obtain expression (39) for the average lifetime

$$\langle\Gamma\rangle = \frac{\Gamma_0}{1+x}. \tag{39}$$

From (31) and (39) we obtain

$$\frac{Z}{Z_\beta} \approx (1+\gamma\Gamma_{0\beta})^{-1} = e^{\ln Z/Z_\beta} = 1+\beta(\bar{u}-u_\beta)+\ln Z/Z_\beta + a_\beta q^2/2 = 1-\gamma\bar{\Gamma}, \quad \bar{\Gamma}=\frac{\Gamma_{0\beta}}{1+\gamma\Gamma_{0\beta}}. \tag{40}$$

Differentiating (40) with respect to β, we obtain

$$\frac{Z}{Z_\beta}(u_\beta-\bar{u}) = \frac{1}{2}q^2\frac{da_\beta}{d\beta} + \beta\frac{\partial(\bar{u}-u_\beta)}{\partial\beta}. \tag{41}$$

Differentiating (37) with respect to γ, we obtain, taking into account the equality $\frac{\partial\langle\Gamma\rangle}{\partial\beta} = \frac{\partial\langle u\rangle}{\partial\gamma}$, that

$$\gamma\langle\Gamma\rangle^2 = \beta\frac{\partial\langle\Gamma\rangle}{\partial\beta} + a_\beta q\frac{dq}{d\gamma}. \tag{42}$$

From (42) and (40)

$$\frac{\partial\langle\Gamma\rangle}{\partial\beta} = \frac{d\Gamma_0}{d\beta}\frac{1}{(1+x)^2}, \quad x=\gamma\Gamma_0, \quad \beta\frac{d\Gamma_0}{d\beta} = \gamma\Gamma_0^2 - a_\beta q\frac{dq}{d\gamma}(1+x)^2. \tag{43}$$

From (31), (40), differentiating (40) with respect to β, we obtain

$$\bar{u}-u_\beta = \gamma\frac{d\Gamma_0}{d\beta}\frac{1}{1+x}, \tag{44}$$

and taking into account (42) we obtain a differential equation for $a_\beta q^2/2$ of the form

$$\gamma(1+x)d(a_\beta q^2/2)/d\gamma - a_\beta q^2/2 = x - \ln(1+x)$$

with the decision

$$a_\beta q^2/2 = \ln(1+x). \tag{45}$$

Then

$$\gamma = \frac{1}{\Gamma_0}(e^{a_\beta q^2/2}-1), \quad 1+x = e^{a_\beta q^2/2}, \quad \langle\Gamma\rangle = \frac{\Gamma_0}{1+x} = \Gamma_0 e^{-a_\beta q^2/2}. \tag{46}$$

Heat fluxes reduce lifetime. From (46) we write the relations for the derivatives

$$a_\beta q\gamma\frac{dq}{d\gamma} = \frac{x}{1+x}, \quad \frac{1}{2}q^2\beta\frac{da_\beta}{d\beta} = -\frac{x}{1+x} = -\gamma\langle\Gamma\rangle. \tag{47}$$

For nonequilibrium temperature (19), taking into account (47), we obtain

$$\theta^{-1} = \beta - \frac{1}{2}q^2\frac{da_\beta}{d\beta}\frac{d\beta}{du_\beta} = \beta + \frac{x}{1+x}\frac{1}{\beta du_\beta/d\beta}. \tag{48}$$

The derivatives of the lifetime from (42), (43), taking into account (47), are equal

$$\beta\frac{d\bar{\Gamma}}{d\beta} = -\frac{\bar{\Gamma}}{1+x}, \quad \beta\frac{d\Gamma_0}{d\beta} = -\Gamma_0, \quad \frac{d\bar{\Gamma}}{d\gamma} = -\langle\Gamma\rangle^2. \tag{49}$$

Using expressions (46), (47), (49) in (48), we find

$$\theta^{-1}\frac{du_\beta}{d\beta} = \beta\frac{du_\beta}{d\beta} + \frac{\gamma\bar{\Gamma}}{\beta}, \quad \theta^{-1} = \beta - \frac{(\bar{u}-u_\beta)}{du_\beta/d\beta} = \beta + \frac{(1-e^{-a_\beta q^2/2})}{\beta du_\beta/d\beta}. \tag{50}$$

For the energy deviation from the equilibrium value, we obtain

$$\bar{u}-u_\beta = -\frac{1}{\beta}\frac{x}{1+x} = -\frac{\gamma\bar{\Gamma}}{\beta} = \frac{1}{\beta}(e^{-a_\beta q^2/2}-1). \tag{51}$$

Integrating expression (49), we obtain

$$\Gamma_0 = \frac{C_\Gamma}{\beta} = C_\Gamma T, \tag{52}$$

where $C_\Gamma$ is independent of $\beta$, $\gamma$, $u_\beta$, $q$. From (49), (51) we find

$$\frac{d\bar{u}}{d\beta} = \frac{du_\beta}{d\beta} + \frac{\gamma \bar{\Gamma}}{\beta^2}(1 + \frac{\bar{\Gamma}}{\Gamma_0}), \quad \frac{d\bar{u}}{d\gamma} = \frac{\bar{\Gamma}}{\beta}(\gamma \bar{\Gamma} - 1). \tag{53}$$

The flows and their derivatives are equal

$$q = [\frac{2}{a_\beta}\ln(1+x)]^{1/2}, \quad \frac{dq}{d\gamma} = \frac{1}{[2a_\beta \ln(1+x)]^{1/2}}\frac{\Gamma_0}{1+x} = \frac{\Gamma_0}{a_\beta q}e^{-a_\beta q^2/2}. \tag{54}$$

Expressions (26), (27) are written in more detail.

In the comments after expression (2), it was noted that the parameter $\gamma$, which is the lifetime conjugate, is associated with the production of entropy. From (46) we obtain

$$\gamma = \frac{1}{\Gamma_0}(e^{a_\beta q^2/2} - 1) \approx \frac{-\Delta s}{\Gamma_0}, \quad \Delta s = s - s_{eq} = -a_\beta q^2/2. \tag{55}$$

Thus, the production of entropy coincides with the parameter $\gamma$ only in the first approximation of the expansion of the exponent in a series. But in (46) and (55), it is precisely the nonlinearity of the quantity $\gamma$ conjugate with the lifetime that is important.

A number of expressions for the nonequilibrium temperature (48), (49), average lifetime (46), and also other expressions for the average lifetime are written down, for example, using expressions (44), (46), (47), (49), (51), we obtain

$$\bar{\Gamma} = \Gamma_0 \frac{\beta(\bar{u} - u_\beta)(1 + q^2 \frac{1}{2}\beta \frac{\partial a_\beta}{\partial \beta})}{q^2 \frac{1}{2}\beta \frac{\partial a_\beta}{\partial \beta}} = -q^2 \frac{1}{2}\beta \frac{\partial a_\beta}{\partial \beta}\frac{1}{\gamma} = a_\beta q \frac{\partial q}{\partial \gamma} = -\Gamma_0 \frac{\beta(\bar{u} - u_\beta)}{(e^{-\Delta s} - 1)}. \tag{56}$$

The obtained relations are generalized to the cases of the presence of other influences on the system, except for the heat flux. For example, if there is still viscous pressure $P^{0v}$, then expression (45) takes the form $a_\beta q^2/2 + (\tau_2/4\eta T)P^{0v}:P^{0v} = \ln(1 + x_1)$, where η is the viscosity.
From (31) and (46) we obtain

$$\langle \Gamma \rangle = \frac{\Gamma_0}{1+x} = \Gamma_0 e^{-a_\beta q^2/2} = \Gamma_0 e^{\Delta s}, \quad \Delta s = -\frac{1}{2}a_\beta q^2 \leq 0. \tag{57}$$

Any flows for one stationary nonequilibrium state near the equilibrium state reduce the lifetime. This applies to substance flows, viscous pressure, and other flows. The same results can be obtained from a comparison with the results of the NSO [7, 8] and informational statistical thermodynamics [5, 6].

This situation can be visualized so that the energy under the influence will more intensively pass through the "potential barrier" to the state of equilibrium from the "potential well".

In this paper, as in [2], the case is considered when $\Delta s = s - s_{eq} = -\Delta \leq 0$. However, for example, in [44] it was shown that in a stationary nonequilibrium state negative entropy fluxes enter the system, and in the general case a situation is possible when $\Delta s > 0$. Then the value $\gamma$ in (55) will be negative, and the value $\langle \Gamma \rangle$ in (57) will be greater than $\Gamma_0$.

Note in conclusion that the approach of the present work is close to [45]: the parameters are set at the level of the macroscopic (phenomenological) description.

# References

1. S. R. DeGroot, P. Mazur, *Non-equilibrium Thermodynamics*. Dover Publications Inc. New York. 1954. 516 p.
2. D. Jou, J. Casas-Vazquez, G.Lebon, *Extended Irreversible Thermodynamics*. Springer, Berlin, Germany, 1993, 442 p.
3. M. A. Leontovich, *Introduction to Thermodynamics. Statistical Physics*, Nauka, Moscow, 1985 (in Russian).
4. U. Seifert, Stochastic thermodynamics, fluctuation theorems, and molecular machines, *Rep. Prog. Phys.* 75 (2012) 126001.
5. R. Luzzi, A. R. Vasconcellos, and J. G. Ramos, *Statistical Foundations of Irreversible Thermodynamics,* Teubner-Bertelsmann-Springer, Stuttgart, Germany, 2000.
6. R. Luzzi, R., A. R. Vasconcellos, J. G. Ramos, et al. Statistical Irreversible Thermodynamics in the Framework of Zubarev's Nonequilibrium Statistical Operator Method. *Theor. Math. Phys.* (2018) 194: 4. https://doi.org/10.1134/
7. D. N. Zubarev, *Nonequilibrium statistical thermodynamics*, Plenum-Consultants Bureau, New York, USA, 1974.
8. D. N. Zubarev, V. Morozov, and G. Röpke, *StatisticalMechanics of Nonequilibrium Processes: Basic Concepts,Kinetic Theory,*Akademie-Wiley VCH, Berlin,Germany, Vol. 1, 1996; D. N. Zubarev, V. Morozov, and G. Röpke, *StatisticalMechanics of Nonequilibrium Processes: Relaxationand Hydrodynamic Processes,*Akademie-Wiley VCH, Berlin, Germany, Vol. 2, 1997.
9. V. V. Ryazanov, Lifetime of system and nonequilibrium statistical operator method. – *Fortschritte der Phusik/Progress of Physics*, 2001, v. 49, N8-9, p.885–893.
10. V. V. Ryazanov, S. G. Shpyrko, 2006 First-passage time: a conception leading to superstatistics. *Condensed Matter Physics*, vol. **9**, No. 1(45), (2006) p. 71-80.
11. V. V. Ryazanov, Lifetime distributions in the methods of non-equilibrium statistical operator and superstatistics. *European Physical Journal B*. Volume 72, number 4, (2009)629–639.
12. V. V. Ryazanov, Nonequilibrium Thermodynamics based on the distributions containing lifetime as thermodynamic parameter. *Journal of Thermodynamics*, Volume 2011, Article ID 203203, 10 pages, 2011. doi:10.1155/2011/203203.
13. V. V. Ryazanov, Nonequilibrium Thermodynamics and Distributions Time to achieve a Given Level of a Stochastic Process for Energy of System. *Journal of Thermodynamics*, vol. 2012, Article ID 318032, 5 pages, 2012. doi:10.1155/2012/318032.
14. V. I. Tikhonov, M. A. Mironov, *Markov processes* M .: Soviet Radio, 1977 (in Russian).
15. J. G. Kirkwood, The statisticalmechanical theory of transport processes I. *J.Chem.Phys.***14**, (1946) 180-201; The statisticalmechanical theory of transport processes II, *J.Chem.Phys.***15**, (1947) 72-76.
16. I. Prigogine, *From Being to Becoming*, Freeman, 1980.
17. R. L. Stratonovich, *Nonlinear Nonequilibrium Thermodynamics*, Springer, Heidelberg, 1992.
18. W. Feller, *An Introduction to Probability Theory and its Applications*, vol.2, J.Wiley, 1971.
19. R. L. Stratonovich, *The elected questions of the fluctuations theory in a radio engineering*, Gordon and Breach, New York, 1967.
20. L. A. Pontryagin, A. A. Andronov, A. A. Vitt, *Zh. Eksp. Teor. Fiz.* 3 (1933) 165 [translated by J. B. Barbour and reproduced in "Noise in Nonlinear Dynamics", 1989, edited by F. Moss and P. V. E. McClintock, Cambridge: Cambridge University Press, Vol. 1, p.329.
21. H. A. Kramers, Brownian Motion on a Field of Force and the Diffusion Model of Chemical Reactions, *Physica* **7** (1940), 284 – 304.
22. N. G. Van Kampen, *Stochastic Processes in Physics and Chemistry*, North-Holland, Amsterdam, 1992.

23. C. W. Gardiner, *Handbook of Stochastic Methods* (2nd edition), Springer, Berlin, 1990.
24. P. Talkner, Mean first passage times and the lifetime of a metastable state. *Z. Phys. B*. **68** (1987) 201–207.
25. V. P. Maslov and M. V. Fedoriuk, *Semi-Classical Approximation in Quantum Mechanics*, Reidel, Dordrecht, 1981.
26. R. S. Maier and D. L. Stein, The escape problem for irreversible systems, *Phys. Rev. E* **48** (1993) 931–938.
27. R. Landauer, Motion out of noisy states, *J. Stat. Phys.* **53** (1988) 233–248.
28. R. G. Littlejohn, The Van Vleck formula, Maslov theory, and phase space geometry, *J. Stat. Phys.* **68** (1992) 7–50.
29. M. V. Day, Recent progress on the small parameter exit problem, *Stochastics* **20** (1987) 121–150.
30. M. I. Dykman, M. M. Millonas, and V. N. Smelyanskiy, Observable and hidden singular features of large fluctuations in nonequilibrium systems, *Phys. Lett. A* **195** (1994) 53–58.
31. R. Dewar, Information theory explanation of the fluctuation theorem in non-equilibrium stationary state, *J.Phys.A: Math. Gen.* **36** (2003), 631-641.
32. Yu. L. Klimontovich, *Statistical Theory of Open Systems*, Springer Science & Business Media, 1994.
33. E. T. Jaynes, Information theory and statistical mechanics. *Phys.Rev*.1957.V.106, n4.P.620-630; Information Theory and Statistical Mechanics II, V.108, n2.P.171-190.
34. J. A. McLennan, *Introduction to Nonequilibrium Statistical Mechanics*, Prentice Hall, New Jersey, 1989.
35. A. I. Olemskoi, A. Ya. Flat, Application of fractals in condensed-matter physics, *Phys. Usp.* **36** (12) 1087–1128 (1993); DOI:10.1070/PU1993v036n12ABEH002208.
36. J. P. Bouchaud, L. F. Gudliandolo, J. Kurchan, *Spin Glasses and Random Fields.*Ed. A. P. Young, Singapore: World Scientific, 1998.
37. S. F. Edwards, *Granular Matter: an Interdisciplinary Approach.* Ed. A. Metha, New York: Springer-Verlag, 1994.
38. A. L. Kuzemsky, Thermodynamic Limit in Statistical Physics, *International Journal of Modern Physics B* 28(9), (2014), 1430004 (28 pages).
39. V. S. Korolyuk and A. F. Turbin, *Mathematical Foundations of the State Lumping of Large Systems*, Kluwer Acad.Publ., 1993.
40. Nihat Ay, Jürgen Jost, HôngVânLê, Lorenz Schwachhöfer, *Information Geometry*. Springer, 2017, 407 pp.
41. J. R. Dorfman, P. Gaspard, Chaotic scattering theory of transport and reaction-rate coefficients, *Phys.Rev. E*. V.51. N1. (1995) P.28-33.
42. P. Gaspard, What is the role of chaotic scattering in irreversible processes? *Chaos*. V.3. N4. (1993) P.427-442.
43. P. Gaspard, J. R. Dorfman, Chaotic scattering theory, thermodynamic formalism and transport coefficients*, Phys.Rev. E*. V.52. N4. (1995) P.3525-3552.
44. I. Prigogine, *Introduction to thermodynamics of irreversible processes*. Springer, USA, 1955.
45. H. Haken, *Information and Self-Organization: A Macroscopic Approach to Complex Systems*, Springer, Berlin, 1988.